\documentclass[12pt]{elsarticle}

\usepackage[utf8]{inputenc}
\usepackage{amsmath}
\usepackage{amsfonts}
\usepackage{amssymb}
\usepackage{graphicx}
\usepackage{subcaption}
\usepackage[]{algorithmic}

\usepackage[]{algorithm2e}
\usepackage{listings}
\usepackage{color}
\definecolor{mygreen}{rgb}{0,0.6,0}
\definecolor{mygray}{rgb}{0.5,0.5,0.5}
\definecolor{mymauve}{rgb}{0.58,0,0.82}

\usepackage{url}
\usepackage{color}
\usepackage{multirow}

\usepackage{pgfplots}
\usepackage{tikz}			
\usetikzlibrary{shapes}

\biboptions{square,sort,comma,numbers}

\newcounter{bla}

\journal{Computer Physics Communications}

\begin{document}
\begin{frontmatter}

\title{Development and performance of a HemeLB GPU code for human-scale blood flow simulation}


\author[a]{I Zacharoudiou}
\author[a]{J.W.S. McCullough}
\author[a,b]{P.V. Coveney \corref{author}}

\cortext[author] {Corresponding author.\\\textit{E-mail address:} p.v.coveney@ucl.ac.uk}
\address[a]{Centre for Computational Science, Department of Chemistry, University College London, UK}
\address[b]{Informatics Institute, University of Amsterdam, Netherlands}

\begin{abstract}
In recent years, it has become increasingly common for high performance computers (HPC) to possess some level of heterogeneous architecture - typically in the form of GPU accelerators. In some machines these are isolated within a dedicated partition, whilst in others they are integral to all compute nodes - often with multiple GPUs per node - and provide the majority of a machine's compute performance. In light of this trend, it is becoming essential that codes deployed on HPC are updated to execute on accelerator hardware. In this paper we introduce a GPU implementation of the 3D blood flow simulation code HemeLB that has been developed using CUDA C++. We demonstrate how taking advantage of NVIDIA GPU hardware can achieve significant performance improvements compared to the equivalent CPU only code on which it has been built whilst retaining the excellent strong scaling characteristics that have been repeatedly demonstrated by the CPU version. With HPC positioned on the brink of the exascale era, we use HemeLB as a motivation to provide a discussion on some of the challenges that many users will face when deploying their own applications on upcoming exascale machines.
\end{abstract}

\end{frontmatter}

{\bf PROGRAM SUMMARY} 

\begin{small}
\noindent
{\em Program Title:} HemeLB (HemePure-GPU)           \\
{\em Developer's repository link:} https://github.com/UCL-CCS/HemePure-GPU \\
{\em Licensing provisions:} LGPLv3  \\
{\em Programming language:} CUDA C++                                  \\

{\em Nature of problem(approx. 50-250 words):} The geometric characteristics of arteries and veins throughout the human body can vary considerably between individuals. This is particularly true for patients with cardiovascular disease. Accurately resolving the dynamics of blood flow within such domains requires a three-dimensional technique that can replicate such variations at high fidelity. The study of full human-scale domains for physiologically relevant timeframes further demands a tool that can be executed efficiently on the advancing technological frameworks available on modern high performance computers. \\ 

{\em Solution method(approx. 50-250 words):} HemeLB uses the lattice Boltzmann method [1,2,3,4] to simulate blood flow in complex, three-dimensional sparse vasculatures. 
A single relaxation time approximation is used [5]. Solid boundaries are modelled using simple bounce-back boundary conditions [6]. Blood flow is driven by applying either velocity or pressure boundary conditions [7].
The localised solution kernels of the algorithm allow for efficient parallelisation of the method to very high core counts. This version of HemeLB outlines the conversion of the code to allow execution on NVIDIA GPUs, currently the most widely used architecture in high performance computers, whilst retaining its capacity for strong scaling to very large core counts.

{\em Additional comments including restrictions and unusual features (approx. 50-250 words):}\\
   \\

\end{small}

\section{Introduction}
In the field of computational biomedicine, significant effort is being invested into the development of the virtual human - a digital twin of an individual's physiology. A full virtual human would allow clinicians, scientists and healthcare professionals to make use of patient-specific simulations and predictive models to optimise the care provided to an individual at all stages of life. Conducting simulations on virtual humans will contribute in a major way to our understanding of the several body systems and processes involved, such as the cardiovascular system and the circulatory physiology. \\
  
Several techniques and models have been developed in this direction, with varying dimensionality; from zero to three dimensional models, each category has its limitations and range of application \cite{shi2011review}. Zero-dimensional or lumped parameter models provide a way of evaluating the haemodynamic interactions among the cardiovascular organs. These models, however, ignore any spatial variation of the fundamental variables, such as pressure and flow, and assume a uniform distribution within the system at a given time. Higher dimensional models on the other hand take into account the spatial variation of these parameters. \\

This is important, for example, for simulating the blood flow through vascular networks, which has been conducted widely using both 1D and 3D solvers \cite{Sheng1995,Qureshi2014,Muller2014,Mynard2015b,Xiao2013,Randles2015,McCullough2021}. Both approaches have advantages and disadvantages relating to implementation, resolution of output, assumptions made, and solution time. For example, 1D models are far simpler to implement and faster to solve, especially as networks get larger; however, the approach makes significant assumptions about the geometric properties of a vessel and the detail of flow patterns obtained is limited. On the other hand, 3D models can generate highly detailed flow fields in patient-specific representations of vessels but do so with greater computational effort and significantly increased quantities of data in/output. A combination of methods is also feasible or desired in some cases. In multi-scale models, low dimensional models are increasingly used to provide the boundary conditions for the more sophisticated, complex and potentially patient-specific 3D models\cite{shi2011review}.\\   

The increasing performance, capacity and availability of computing architectures means that 3D models of large and detailed geometries can be studied in a tractable time frame. This is further enhanced when the 3D code scales efficiently up to large core counts. 
This is extremely useful and of paramount importance, as conducting simulations on virtual humans will demand codes that can both execute and scale efficiently on large-scale supercomputers.\\

The purpose of this paper is to provide another step forward in the effort to make the virtual human a reality by taking advantage of the accelerators such as Graphics Processing Units (GPUs) that are becoming commonplace on supercomputers globally. We do this by developing a GPU version of HemeLB \cite{Mazzeo2008,Groen2013}, a high-performance lattice Boltzmann (LB) based fluid flow solver for simulating blood flow on patient specific images obtained from medical scans. HemeLB has been optimized for the sparse geometries characteristic of vascular trees and has demonstrated strong scaling on hundreds of thousands of CPU cores on non-accelerated HPC including BlueWaters and SuperMUC-NG. Our vision for this work is to demonstrate HemeLB’s capacity for execution on the largest and fastest current generation machines and prepare it for upcoming exascale machines. In both of these settings, a considerable part of the performance is sourced from the presence of accelerators on the nodes, typically GPUs. Here we present our implementation of a GPU version of HemeLB and report performance and scaling analysis up to tens of thousands of GPUs.\\

\section{Numerical Methods}
In this paper we use the lattice Boltzmann method (LBM) to solve 3D blood flow through a vascular geometry. A brief outline of this approach is presented in this section.

\subsection{Equations of Motion}
Modeling 3D blood flow is a problem of computational fluid dynamics.  
The blood consists primarily of blood cells suspended in blood plasma. From a rheological perspective, blood plasma has been considered a Newtonian fluid, while whole blood behaves as a non-Newtonian fluid; whole blood, however, follows Newtonian behavior when the shear rate exceeds 100 $s^{-1}$. 
Therefore, for flows in large blood vessels, e.g. aorta or large arteries, the effect of the non-Newtonian nature of blood is not significant and considering the flow as Newtonian is a satisfactory assumption.         

The hydrodynamic incompressible equations of motion that describe the blood flow are the continuity, eq.~\ref{eq:continuity}, and the Navier-Stokes, eq.~\ref{eq:Navier-Stokes}, equations \cite{Xiao2013,shi2011review, figueroa2017blood}
\begin{equation}\label{eq:continuity}
\partial_{t}\rho + \partial_{\alpha}(\rho u_{\alpha})=0\;,
\end{equation}
\begin{equation}\label{eq:Navier-Stokes}
\partial_{t}(\rho u_{\alpha}) + \partial_{\beta}(\rho u_{\alpha}u_{\beta}) =-\partial_{\alpha}p+\partial_{\beta}\left[\eta\left(\partial_{\beta}u_{\alpha}+\partial_{\alpha}u_{\beta}\right) \right]\;,
\end{equation}
where $\bf{u}$, $\rho$, $p$, $\eta$ are the fluid velocity, density, pressure and dynamic viscosity respectively.

\subsection{The lattice Boltzmann method}
It is not the purpose of this paper to give a full and detailed description of the LBM, as this is widely available in the literature e.g.\ \cite{Succi2001,Mohamad2011,Guo2013,Kruger2017,Succi2018}. We will, however, summarise the key features of our implementation.\\

To solve the hydrodynamic equations, eqs.~\ref{eq:continuity}, \ref{eq:Navier-Stokes}, with the LBM, the domain is partitioned into a Cartesian grid with a constant spacing of $\Delta x$ in all 3D directions. At each nodal location, $\textbf{x}$, a discrete set of distribution functions, $f_i(\textbf{x},t)$, is assigned to represent the amount of fluid moving in direction $i$ at time $t$. 
In this work, we use a three-dimensional model with 19 discrete velocity vectors (D3Q19), where fluid can stay at the current location or move to one of the 18 neighbours described by the sets: $i = 1-6$ $\left[ \left(\pm 1, 0, 0 \right),\left(0, \pm 1, 0 \right),\left(0, 0, \pm 1 \right) \right)$ and $i = 7-18$ $\left[\left(\pm 1, \pm 1, 0 \right),\left(\pm 1, 0, \pm 1 \right),\left(0, \pm 1, \pm 1 \right) \right]$. The flow described by $f_i(\textbf{x},t)$ evolves over the time step $\Delta t$

\begin{subequations}\label{eq:evolution_distr_funct}
\begin{align}
\mbox{Collision step:} \qquad& f^{\prime}_i({\bf x},t)=f_i({\bf x},t) -\frac{\Delta t}{\tau}(f_i(\textbf{x},t)-f_i^{eq}(\textbf{x}, t)) \;, \label{eq:evolution_eq_fi_col}\\
\mbox{Streaming step:} \qquad & f_i({\bf x}+{\bf c_i} \Delta t,t + \Delta t)= f^{\prime}_i({\bf x},t) \;.\label{eq:evolution_eq_fi_prop}
\end{align}
\end{subequations}

\noindent Eq.~\ref{eq:evolution_distr_funct} states that the time evolution of the distribution functions proceeds in two steps: (a) a collision step with $f_i$ relaxing towards their equilibrium state $f_i^{eq}$ (Maxwell-Boltzmann distribution) over a timescale $\tau$, within a single relaxation time approximation \cite{bhatnagar1954model}, giving locally the updated or post-collisional distribution functions $f^{\prime}_i$ and (b) a streaming step with velocity $\textbf{c}_i$ to the neighbouring lattice point ${\bf x}+{\bf c_i} \Delta t$ at the next time step $t+\Delta t$.  \\

The equilibrium distributions functions are defined as a power series in the velocity
\begin{equation}
\label{eq:feq}
f_i^{eq}(\textbf{x}, t) = w_i \rho(\textbf{x},t) \left(1 + \frac{\textbf{c}_i\cdot \textbf{u}}{C_s^2} + \frac{(\textbf{c}_i\cdot \textbf{u})^2}{C_s^4} - \frac{|\textbf{u}|^2 }{C_s^2}\right),
\end{equation}
with the coefficients, $w_i$, for the D3Q19 model, being 1/3 for $i = 0$ (the source node), 1/18 for $i = 1-6$ and 1/36 for $i = 7-18$. 
$C_s$ represents the speed of sound of the fluid and evaluates to $\frac{1}{\sqrt{3}}$. 

A Chapmann-Enskog expansion \cite{Luo_2000_PhysRevE.62.4982} can be used to demonstrate that this framework leads to the hydrodynamic equations, continuity eq.~\ref{eq:continuity} and Navier-Stokes eqs.~\ref{eq:Navier-Stokes} in a low Mach number limit. 

Local macroscopic properties of density and momentum can be determined from moments of the $f_i(\textbf{x},t)$ population. In the absence of forces these are given by

\begin{equation}
\label{eq:Density}
\rho(\textbf{x},t) = \sum_{i}f_i(\textbf{x}, t) = \sum_{i}f_i^{eq}(\textbf{x}, t),
\end{equation}
\noindent and,
\begin{equation}
\label{eq:Momentum}
\rho(\textbf{x},t)\textbf{u} = \sum_{i}f_i(\textbf{x},t)\textbf{c}_i = \sum_{i}f_i^{eq}(\textbf{x},t)\textbf{c}_i,
\end{equation}

\noindent respectively. Other relevant physical properties of pressure,
\begin{equation}
\label{eq:Pressure}
p(\textbf{x},t) = C_s^2 \rho(\textbf{x}, t),
\end{equation}
\noindent and viscosity,
\begin{equation}
\label{eq:Viscosity}
\eta = \rho C_s^2 \left(\tau-\frac{\Delta t}{2}\right),
\end{equation}
\noindent arise from the Chapmann-Enskog expansion process.

\subsection{Boundary conditions}
We encounter two different types of boundary conditions (BCs) within the lattice Boltzmann method: a) the BCs at solid surfaces and b) the BCs at the inlets - outlets of the simulation domain. The former refers to enforcing the no-slip boundary condition on the velocity field, while the later on how to drive the fluid flow. 
Within the lattice Boltzmann framework these aim to establish certain conditions at a given boundary site by explicitly determining the values of the unknown post-collisional distribution functions propagating from unknown locations outside the geometry into the simulation domain after the streaming step, see eq.~\ref{eq:evolution_eq_fi_prop}.  

\subsubsection{No-slip boundary conditions}
We enforce the no-slip boundary condition at solid surfaces (blood vessels' walls) by applying the mid-link bounce-back rule proposed by \citet{ladd2001lattice}. Populations streaming towards solid nodes are reflected back towards the lattice nodes they came from, resulting in recovering the wall location (zero velocity) approximately half-way between the fluid and solid node. 

\subsubsection{Inlet - Outlet boundary conditions}
The blood flow is driven by applying either velocity or pressure BCs at the inlets and outlets, termed {\it iolets}, of the domain. 
BCs can be set independently for the two types of {\it iolets} encountered. 
Pressure BCs are imposed by assigning a target pressure at the boundary fluid nodes, which for the LBM is achieved through assigning the value of the local density ({\it ghost} density). 
For velocity BCs, we follow Ladd's method \cite{ladd_1994} to impose the expected velocity profile (e.g. parabolic, Womersley). This is based on a modification of the bounce-back scheme with a correction term $-2w_i\rho {\bf u} \cdot {\bf c_i} / C_s^2$. A more detailed description on the implementation of BCs within HemeLB is provided by \citet{Nash2014}. 

\section{GPU code implementation}
In this section we give an overview of the GPU code implementation. First we provide some general background information on the LB algorithm and the HemeLB code. Then we describe the steps taken to port the HemeLB code to GPU  architecture. 


\subsection{General background}\label{general_background}
The lattice Boltzmann algorithm	proceeds in the following way: 
\begin{enumerate}
\item Initialise macroscopic quantities, density $\rho$ and velocity ${\bf u}$, required for the initialisation of the distribution functions $f_i$ $(i=0-18)$ to their equilibrium value $f_i^{eq}$ using eq.~\ref{eq:feq}.
\item Collision step: evaluate the updated distribution functions $f^{\prime}_i({\bf x},t)$ according to eq.~\ref{eq:evolution_eq_fi_col}.  
\item Streaming step: the updated distribution functions $f^{\prime}_i({\bf x},t)$ propagate to the neighbouring fluid site, ${\bf x}+{\bf c_i}\Delta t$, see eq.~\ref{eq:evolution_eq_fi_prop}. 
\item Apply BCs: solid-fluid BCs and Inlet/Outlet BCs.
\item Repeat steps 2-4 timeSteps-times. These steps represent the core of the LBM algorithm. 
\end{enumerate} 

The above scheme with the collision step preceding the streaming step is refered to as the Push-scheme \cite{wellein2006single}. It would also be possible to have the streaming step first, by gathering the distribution functions from the neighbouring fluid sites to a given fluid site and then perform the collision step. This is known as the Pull-scheme \cite{wellein2006single,tran2017performance}. Further discussion on the above 2 schemes will follow in section~\ref{sec:Future_plans}.

The LBM is inherently amenable to parallelisation owing to its local nature and thus presents itself as a candidate for extreme parallelism on modern supercomputers.  
Exchange of data between neighbouring MPI ranks requires only nearest neighbour information and takes place during the streaming step, when the updated distribution functions stream in and out of the domain assigned to each MPI rank. 
The fluid sites of these shared edges are labeled as {\it domain-edge} sites, while the ones not requiring any exchange of information with neighbouring MPI ranks are labeled as {\it mid-domain} sites.   
Hence, a highly effective way of hiding the MPI communication overhead and enabling scaling of LBM algorithms up to extreme scales is by performing the following at each LBM iteration:

\begin{enumerate}
\item First, perform collision - streaming at {\it domain-edge} sites.
\item Then, issue the MPI exchange for the {\it domain-edge} sites.  
\item Finally, perform collision - streaming at {\it mid-domain} sites, while overlapping these computations with the MPI data exchange.   
\end{enumerate}

\noindent  
HemeLB follows the same approach and registers the following steps to be executed at each iteration (timeStep) through the $\it StepManager$ object
\begin{enumerate}
\item PreSend: Collision - streaming at {\it domain-edge} sites. 
\item Send: Issue the MPI exchange of the updated distribution functions involved at {\it domain-edge} sites
\item PreReceive: Collision - streaming at {\it mid-domain} sites.
\item Receive: Wait for the MPI exchange of the updated distribution functions to complete. 
\item PostReceive: Place the received distribution functions in the appropriate streaming location. 
\item EndIteration: Perform necessary calculations at the end of the LB iteration for updating the property cache, containing macrovariables of interest, e.g. density, velocity etc. Finally, swap the distribution functions $f^{old}$ (pre-collision) and $f^{new}$ (post-collision).       
\end{enumerate}

\subsection{Porting the HemeLB code to GPU architecture} 
Here we describe the steps taken to port the single component HemeLB-CPU code to GPU architecture. 
HemeLB-CPU is written in C++. 
The GPU version of HemeLB has been developed using the CUDA computing platform (CUDA C++) to run on NVIDIA's GPUs. 
We must note that, as we are moving into an era where there may be more kinds of GPUs available (NVIDIA, AMD, INTEL), we would like to make the code eventually platform agnostic.  
To this direction, recent development efforts were carried out porting the  CUDA code to the HIP runtime API, making it able to run on AMD GPUs as well.  
Here, however, we will restrict ourselves to the CUDA version of the code. 

Given the complex nature of the existing CPU version of HemeLB, our initial attempt for porting the HemeLB code on GPU architectures focused on exporting the compute intensive parts onto the GPU (device), without making significant changes to the remaining structure of the code.

The steps taken involve the following:
\begin{enumerate}
\item Initialise the GPU.
\item Implement the GPU collision - streaming kernels.
\item Arrange the exchange of data between the device (GPU) and the host (CPU), i.e. the memory copies from the device to host (D2H) and from the host to device (H2D).
\item Arrange the CUDA streams for the various GPU operations and the appropriate synchronisation points ({\it cudaStreamSynchronize}).   
\end{enumerate}

\subsubsection{Initialisation of the GPU}
Initialising the GPU takes place at the beginning of the simulation and involves allocating memory and copying the data that will reside on the GPU for the duration of the simulation. 
More specifically this involves:    
\begin{itemize}
\item Allocate memory on the GPU global memory and perform a H2D memory copy for the distribution functions ($f^{old}$, $f^{new}$) and the streaming indices; the later refer to the GPU memory locations that the post-collision populations will stream to. A change of the data layout is performed for these data structures. More details on this are provided in section ~\ref{optimisation_strat}. 
\item Allocate memory (GPU global memory) and perform a H2D memory copy for the following:
\begin{enumerate}
\item information for boundary links' intersections (wall/iolet)
\item information for iolets: 
\begin{enumerate}
\item total number of iolets on local MPI rank
\item fluid sites' range associated with each unique iolet
\item normal vector at each iolet 
\end{enumerate} 
\item streaming indices for the received distribution functions $f^{new}$ at shared edges 
\end{enumerate}
\item Allocate memory (GPU global memory) related to BCs at iolets:
\begin{enumerate}
\item for the case of Pressure BCs the {\it ghost} density for each iolet.
\item for the case of Velocity BCs the correction term for each iolet fluid site.
\end{enumerate}
\item Copy to GPU constant memory the following:
\begin{enumerate}
\item discrete velocity directions ${\bf c_i}$ of the LB model (D3Q19).
\item inverse velocity directions (related to the bounce-back scheme).
\item weights $w_i$ for calculating $f^{eq}$.
\item relaxation time $\tau$ and its inverse.
\end{enumerate}

\end{itemize}

\subsubsection{Collision - streaming GPU kernels}
The compute intensive parts of the code that were exported on the GPU involve mainly the collision - streaming steps of the LB algorithm. 
HemeLB distinguishes the following 6 types of collision – streaming, depending on the type of fluid sites and their corresponding streaming links: 1) Inner domain: only fluid sites without any links to any type of boundaries (walls/iolets), 2) Walls: fluid sites with a link to a solid surface, 3) Inlet, 4) Outlet, 5) Inlet with Walls and 6) Outlet with Walls. Hence, GPU CUDA kernels were initially implemented to account for each one for the above collision – streaming types.
We must note that we have eventually merged the first 2 types of collision - streaming kernels, which provided a small gain in performance. 
All collision - streaming kernels can be potentially launched within the {\it PreSend} ({\it domain-edge} sites) and the {\it PreReceive} ({\it mid-domain} sites) steps of the code.

\begin{figure}[!ht]
   \centering
   \includegraphics[width=0.98\textwidth]{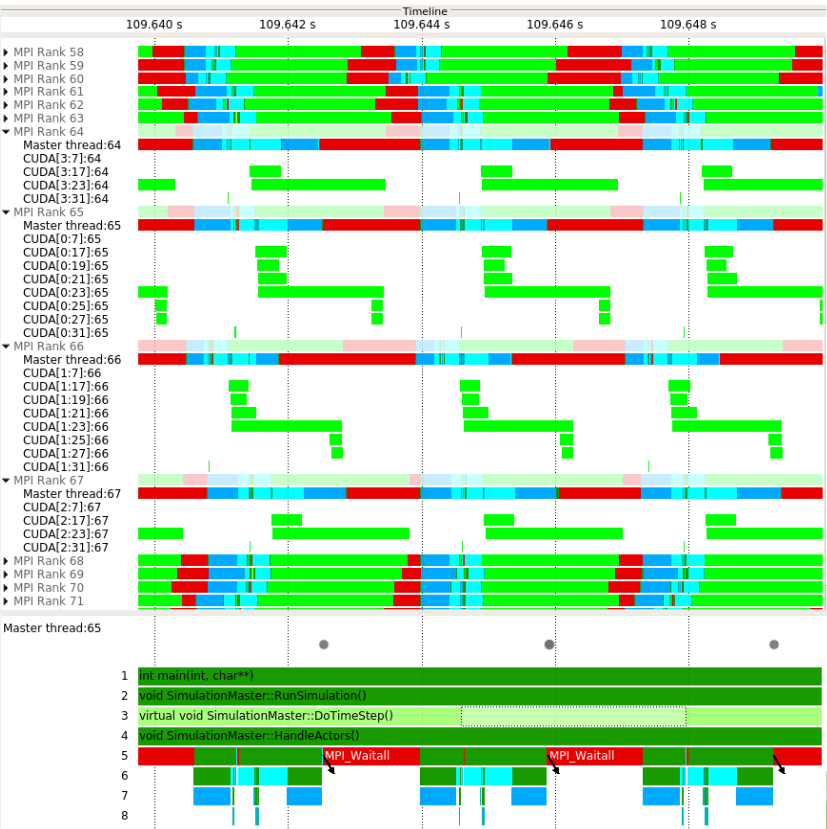}
   \caption{HemeLB-GPU execution timeline: Focus of analysis zoom (3 timesteps) showing GPU CUDA kernels running concurrently in different CUDA streams for MPI ranks 64 to 67. Results from simulation on Juwels Cluster (Julich Supercomputing Centre) with 32 nodes and 129 MPI processes driving 128 Tesla V100 GPUs \cite{Wylie2020GPU}.    
   } 
   \label{fig:profile_kernel_overlap}
\end{figure}

HemeLB groups the fluid sites in a consecutive ascending order depending on their collision - streaming type, with the corresponding fluid sites' range passed as an argument to the appropriate CUDA kernel.  
When the CUDA kernels are launched, their threads (bundled into warps) are more likely to execute the same code branches, avoiding consequently any control divergence. 
Moreover, these kernels reside on different CUDA streams, see Fig.~\ref{fig:profile_kernel_overlap}; hence, they can run concurrently and offer further acceleration of the computations. This is in contrast to the HemeLB-CPU code where the corresponding components execute in a serial manner. 

We implemented the collision - streaming GPU CUDA kernels with the following assumptions: 
\begin{enumerate}
\item Lattice model: D3Q19.
\item Collision operator: Single relaxation time approximation, i.e. BGK collision operator \cite{bhatnagar1954model}.
\item Wall boundary conditions: Mid-link bounce-back \cite{ladd2001lattice}.
\item Inlet/Outlet boundary conditions: Any type of inlet and outlet BCs supported by HemeLB, i.e. pressure or velocity BCs, to drive the blood flow. 
\end{enumerate}
\noindent 
Each CUDA kernel performs the collision and streaming step, as well as applies the appropriate BCs when applicable, by determining the unknown populations $f^{new}$ after the streaming step.

Each thread in the CUDA kernels performs the following:
\begin{enumerate}
\item loads the $f^{old}$ distribution functions from the GPU global memory into local memory.
\item calculates the hydrodynamic macrovariables ($\rho$, ${\bf u}$), using eqs.~\ref{eq:Density},~\ref{eq:Momentum}.
\item calculates the equilibrium $f^{eq}$ values (eq.~\ref{eq:feq}).
\item calculates the post-collision distribution functions locally (eq.~\ref{eq:evolution_eq_fi_col}).
\item performs the streaming step (eq.~\ref{eq:evolution_eq_fi_prop}). If the thread encounters any boundary links during streaming (wall/iolet), apply the appropriate BCs to determine the unkonwn populations $f^{new}$. 
\item stores the updated values $f^{new}$ into the GPU global memory.
\item stores the hydrodynamic macrovariables ($\rho$, ${\bf u}$) into the GPU global memory at a specified frequency, e.g. every 100 time-steps. 
\end{enumerate}

\begin{figure}[!ht]
   \centering
   \includegraphics[width=0.98\textwidth]{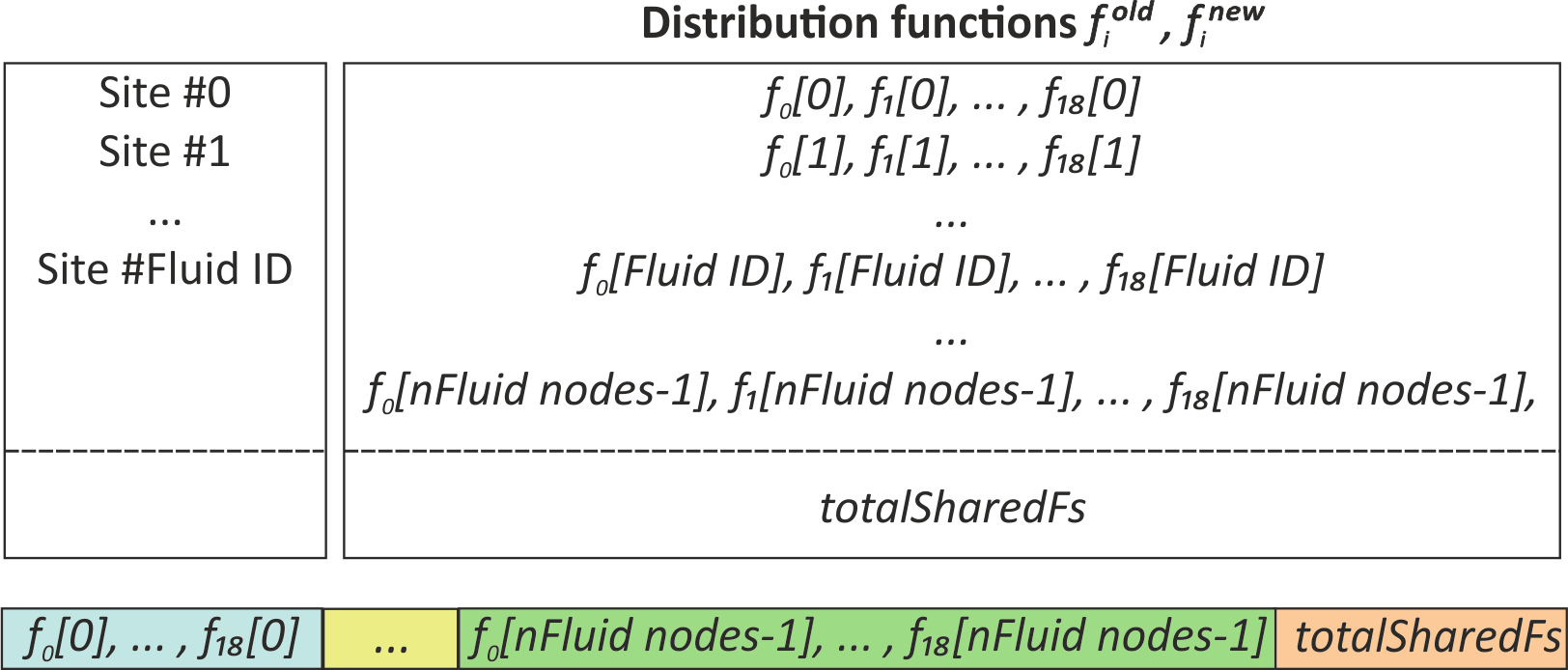}
   \caption{Array of Structures scheme (HemeLB-CPU version). Distribution functions $f^{old}$ (pre-collision) and $f^{new}$ (post-collision) are arranged based on the fluid site ID in a 1D array.   
   } 
   \label{AoS}
\end{figure}

\begin{figure}[!ht]
   \centering
   \includegraphics[width=0.98\textwidth]{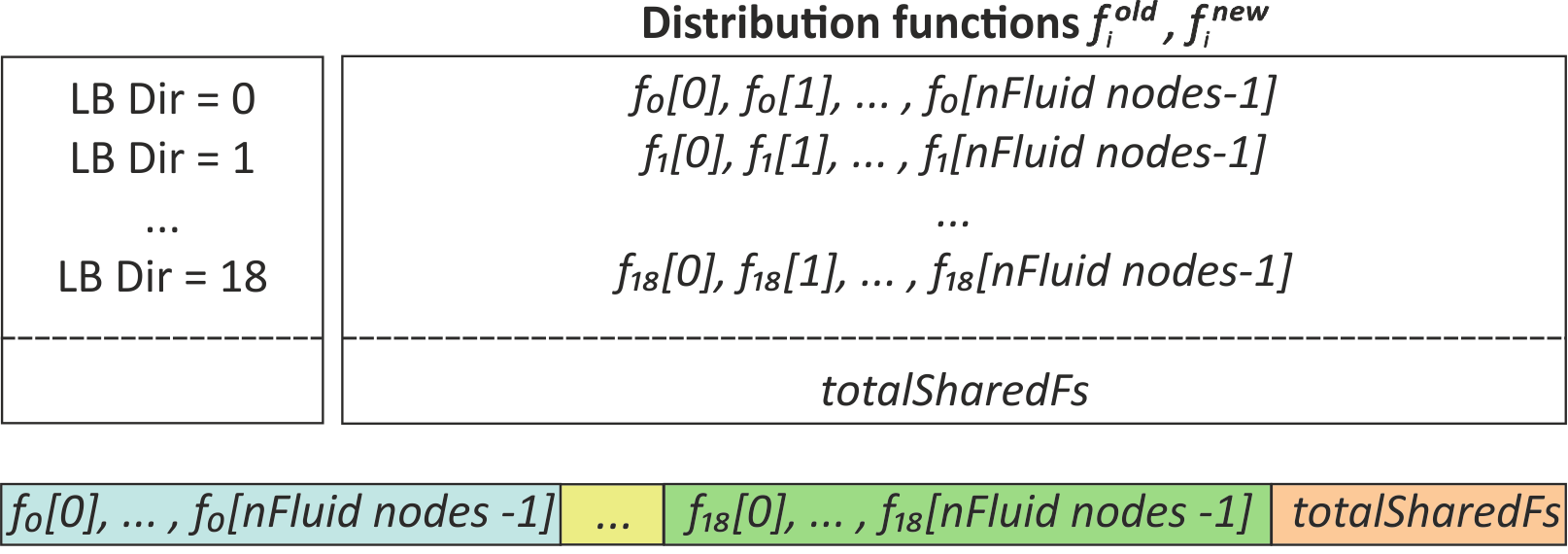}
   \caption{Structure of Arrays scheme (HemeLB-GPU version). Distribution functions $f^{old}$ (pre-collision) and $f^{new}$ (post-collision) are arranged based on the discrete velocity directions of the lattice model (D3Q19) in a 1D array. 
   }
   \label{SoA}
\end{figure}

Populations that stream out of the simulation domain at domain edges during the streaming step are appended at the end of the $f^{new}$ 1D array in {\it totalSharedFs}, see Figs.~\ref{AoS}-\ref{SoA}. 
The collision - streaming CUDA kernels for these sites are launched first during the step {\it PreSend}, as discussed in section ~\ref{general_background}. 
A CUDA synchronisation barrier ({\it cudaStreamSynchronize}) ensures that all GPU collision - streaming kernels in {\it PreSend} have completed their task. A D2H asynchronous memory copy ({\it cudaMemcpyAsync}) is then issued to transfer the populations in {\it totalSharedFs} to the host and arrange the MPI send to the appropriate neighbouring rank (step {\it Send}). 
Once transfered on the receiving MPI rank, a H2D asynch. mem. copy is issued and the data are placed in the {\it totalSharedFs} location of the $f^{old}$ 1D array (({\it PostReceive} step). 
Finally, a GPU kernel ({\it GPU\_StreamReceivedDistr}) performs the appropriate re-allocation of the received distribution functions, from the {\it totalSharedFs} in $f^{old}$ into the destination buffer $f^{new}$. 

Once the above re-allocation 
is complete and all collision - streaming GPU kernels in step {\it PreReceive} ({\it mid-domain } sites) have also completed their task (use {\it cudaStreamSynchronize}), swapping of the populations takes place in step {\it EndIteration}; this ends the LB algorithm for the current time-step.  

\subsubsection{Optimisation strategies}\label{optimisation_strat}
\begin{figure}[!ht]
   \centering
   \includegraphics[width=0.98\textwidth]{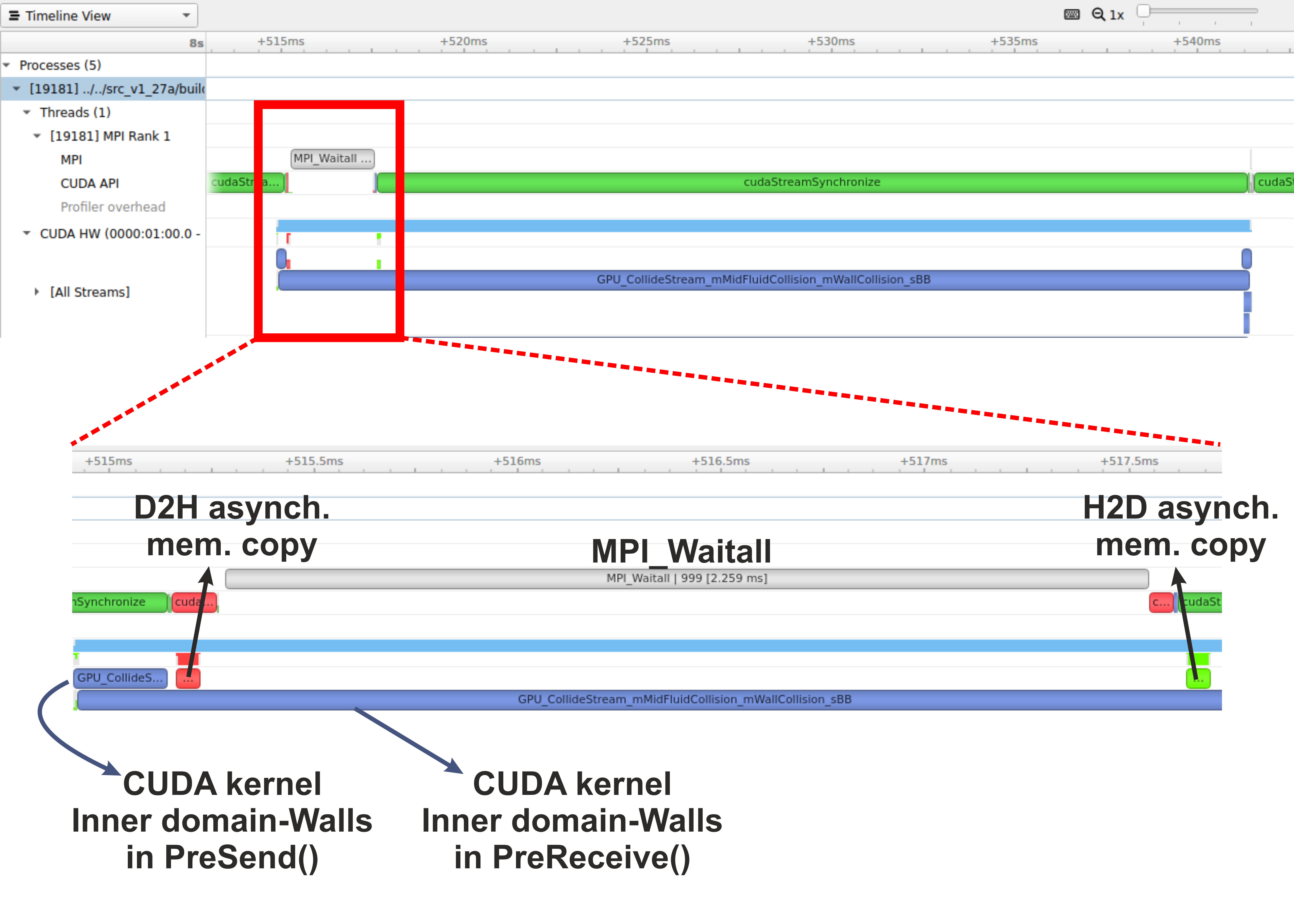}
   \caption{Profiling the HemeLB-GPU code using NSight Systems. Focus of analysis zoom (1 timestep) showing GPU CUDA kernels (blue) in different streams, as well as the asynchronous memory copies (D2H and H2D). The D2H memory copy sends the data to the host to be sent to neighbouring MPI ranks; once the MPI send is complete and the updated distribution functions have been received on the host, the H2D mem. copy transfers these to the device to be placed in {\it totalSharedFs}.  
   }
   \label{fig:Profiling_kernels_memCopies}
\end{figure}

The main optimisation strategies used during the GPU code development were:
\begin{itemize} 
\item {\bf Change of data layout}. 
HemeLB stores the distribution functions ($f^{old}$, $f^{new}$) as an {\it Array-of-Structures} (AoS), where data is arranged in a 1D array in system memory based on the fluid site index, see Fig.~\ref{AoS}. The GPU version stores these data following the {\it Structure-of-Arrays} (SoA) scheme; data is arranged based on the LB discrete velocity directions, see Fig.~\ref{SoA}. SoA scheme is more suitable for the GPU architecture as demonstrated by \citet{tran2017performance}.      
Finally the array appended at the end of the 1D arrays, labeled as {\it totalSharedFs}, that corresponds to the buffer for placing and receiving the post-collision distribution functions at shared edges remains the same as in the CPU version.

\item {\bf Change the sequence of steps} registered in {\it StepManager}, compared to HemeLB-CPU. HemeLB-GPU reorders the pattern of MPI exchanges of data. 
Instead of ordering the steps as {\it PreSend} $\rightarrow$ {\it Send} $\rightarrow$ {\it PreReceive}, the sequence is modified to {\it PreSend} $\rightarrow$ {\it PreReceive} $\rightarrow$ {\it Send}.
Effectively, all GPU CUDA collision - streaming kernels ({\it PreSend}:{\it domain-edge}  and {\it PreReceive}:{\it mid-domain}) are launched first, with control returned to the host, before issuing the MPI exchange of data. This allows a better overlap of the computations and improves consequently the code's performance.

\item {\bf Use of different cuda streams for all GPU operations.} 
This allows overlapping the CUDA kernels' execution, see Fig.~\ref{fig:profile_kernel_overlap}, and the asynchronous memory copies, Device to Host (D2H) and Host to Device (H2D), see Fig.~\ref{fig:Profiling_kernels_memCopies}. 
 
\item {\bf Swap the distribution functions at the end of the LB iteration.} Exchange the pointers for the fundamental LB data ($f^{old}$, $f^{new}$), instead of explicitly swapping the data. 

\end{itemize}

\subsection{Compiling - Running a simulation} \label{sec:compiling}
Compiling the code is a two-stage process. First, the user must build the dependencies, before compiling the source code to generate the executable {\it hemepure\_gpu}. 
A full build script that can be used to compile the code is provided in Listing~\ref{list:Full_BuildScript}. 
The modules that need to be loaded (line 1) include the modules for the compiler (e.g. GCC), CUDA, MPI and CMake. 
The user should further modify accordingly: a) the lines refering to the boundary conditions (iolets/walls: lines 31-34) and b) the GPU compute capability of the NVIDIA GPUs available on the system (line 35). 
For pressure BCs the user should specify {\it NASHZEROTHORDERPRESSUREIOLET} and for velocity BCs {\it LADDIOLET}.

\begin{lstlisting}[caption={Full build script to first build the dependencies and then compile the source code. 
}, label={list:Full_BuildScript}, 
language=bash]
## MODULE loads
export CC=$(which gcc)
export CXX=$(which g++)
export MPI_C_COMPILER=$(which mpicc)
export MPI_CXX_COMPILER=$(which mpicxx)

export SOURCE_DIR=/Path_To_GPU_code

## HEMELB build
# 1) Dependencies
BuildDep(){
cd $SOURCE_DIR/dep
rm -rf build
mkdir build
cd build
cmake -DCMAKE_C_COMPILER=${CC} -DCMAKE_CXX_COMPILER=${CXX} ..
make -j  && echo "Done HemeLB Dependencies"
cd ../..
}

# 2) Source code
BuildSource(){
cd $SOURCE_DIR/src
rm -rf build
mkdir build
cd build
cmake  \
  -DCMAKE_CXX_FLAGS="-std=c++11 -g -Wno-narrowing" \
  -DCMAKE_C_COMPILER=${CC} \
  -DCMAKE_CXX_COMPILER=${CXX} \
  -DHEMELB_INLET_BOUNDARY="NASHZEROTHORDERPRESSUREIOLET"\
  -DHEMELB_WALL_INLET_BOUNDARY="NASHZEROTHORDERPRESSURESBB"\
  -DHEMELB_OUTLET_BOUNDARY="NASHZEROTHORDERPRESSUREIOLET"\
  -DHEMELB_WALL_OUTLET_BOUNDARY="NASHZEROTHORDERPRESSURESBB"\
  -DCMAKE_CUDA_FLAGS=``-ccbin g++ -gencode arch=compute_70, code=sm_70 -lineinfo --ptxas-options=-v --disable-warnings" \ 
  	$SOURCE_DIR/src				    
make -j && echo "Done HemeLB Source"
cd ../..
}

BuildDep
BuildSource
echo "Done build all"

\end{lstlisting}
As long as the compiler remains unchanged, the dependencies only need to be built once. Hence, BuildDep (line 41) can be commented out for subsequent  compilations with different options. 

Running the executable can be done in the same way as the CPU version;
 a detailed description of the input file and how to run a simulation is provided in the official HemeLB website \cite{hemelb_2018}.

\section{Large scale performance comparison}
\begin{figure}[!ht]
 \begin{subfigure}{0.3\textwidth}
   \centering
   \includegraphics[width=0.98\textwidth]{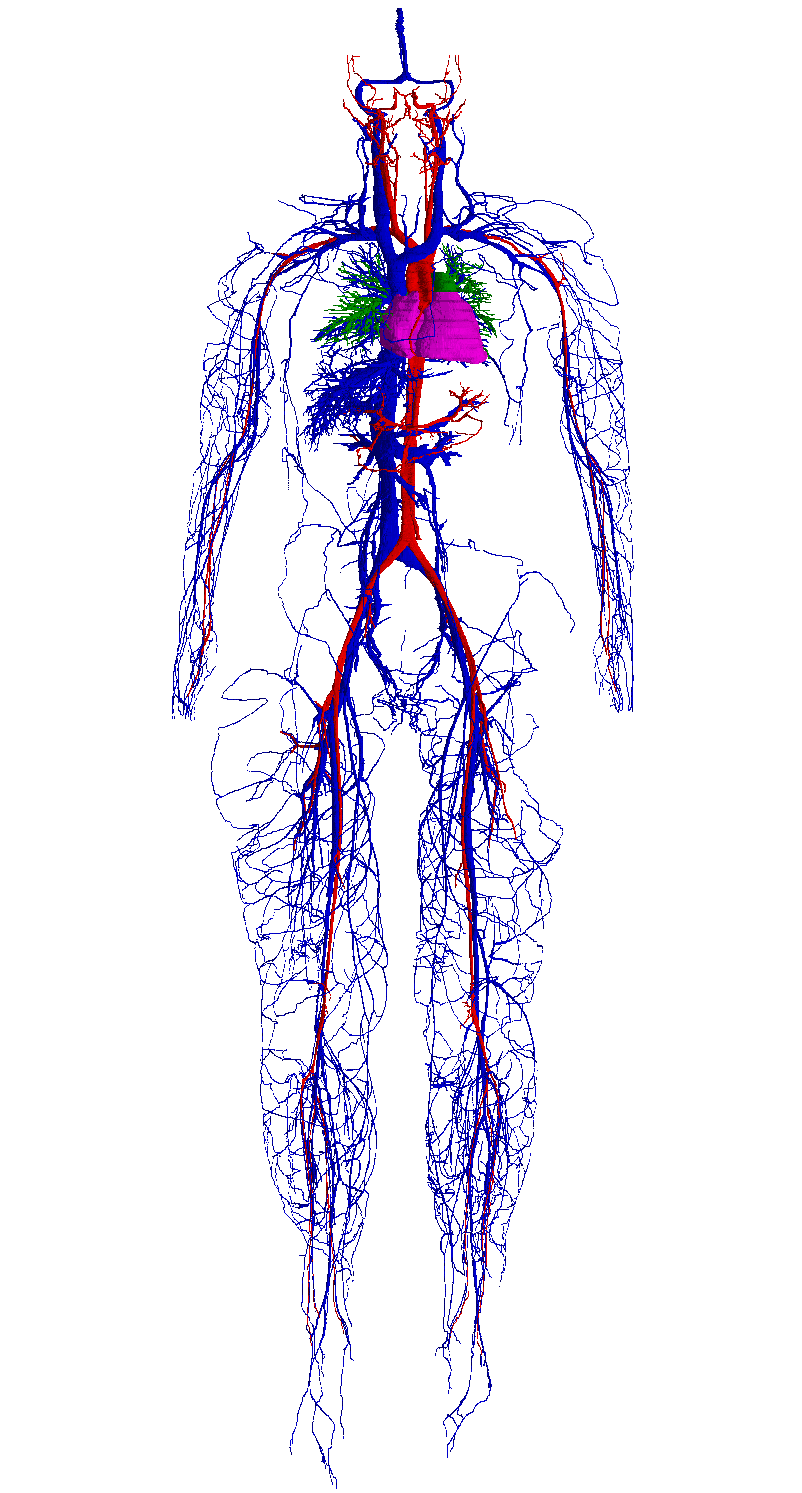}
   \caption{Full human vessels}
   \label{FHVdomain}
 \end{subfigure}
  \begin{subfigure}{0.3\textwidth}
   \centering
   \includegraphics[width=0.98\textwidth]{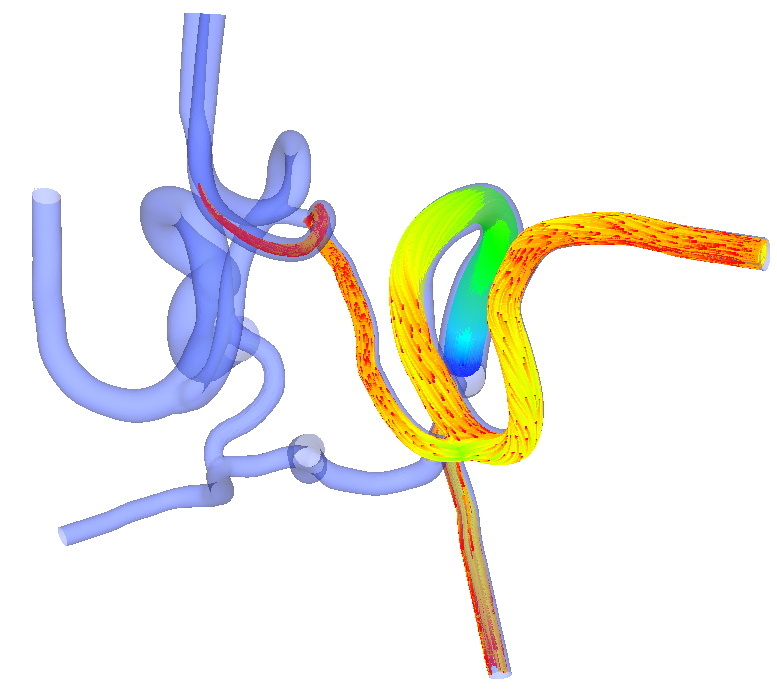}
   \caption{Circle of Willis}
   \label{COWdomain}
 \end{subfigure}
  \begin{subfigure}{0.3\textwidth}
   \centering
   \includegraphics[width=0.98\textwidth]{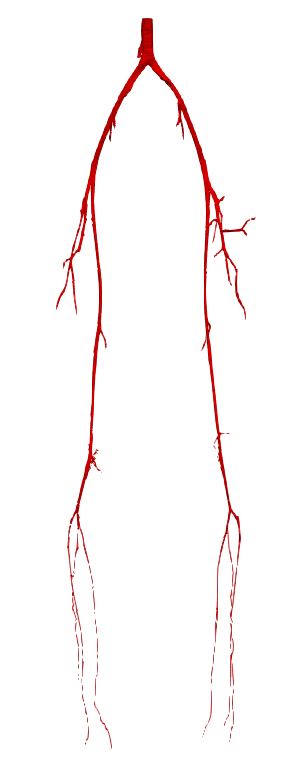}
   \caption{Arteries of the legs}
   \label{ALdomain}
 \end{subfigure}
  \caption{Illustrations of the test domains used to test the large-scale performance of HemeLB's GPU and CPU versions. The blue veins from (a) were used as a smaller case in terms of lattice sites. The arterial legs (c) was a $66.4\times10^6$ 
  site domain used on a machine that posed particular restrictions on memory available for job execution.}
  \label{fig:TestDomains}
\end{figure}

HemeLB has been specifically optimized to allow excellent strong scaling performance on the sparse and irregular geometries that are characteristic of vascular domains. In our comparisons of performance on CPU and GPU based architectures, we consider two particular domains of different resolutions. The first is a discretization of the full human venous tree consisting of approximately $1.6\times10^9$ 
lattice sites, whilst the second represents the circle of Willis arterial structure found in the brain possessing $10.2\times10^9$ 
sites. Whilst both of these domains are of significant magnitude, we demonstrate that even these are not sufficient to adequately occupy current petascale machines to their full extent. These domains are illustrated in Figure \ref{fig:TestDomains}.

In this study, we are focussed on the strong scaling performance of HemeLB. Because of the fixed and irregular shape of vascular geometries we are considering, it is difficult to accurately partition the domain to ensure an even balance of work between processors necessary for a rigorous weak scaling analysis. Assessing and comparing strong scaling performance between CPUs and GPUs is not a straightforward task due to the inherent differences between their execution. Adding to this are the differences in how HPC facilities have packaged resources onto a node of their machines. In Table \ref{tab:PerformanceMetrics} we identify potential metrics for comparing the large-scale performance of CPU and GPU codes both in terms of scale (x-axis) and performance (y-axis) with their benefits and drawbacks.

\begin{table}
\caption{Various potential metrics for assessing the strong scaling performance of CPU and GPU codes}
\begin{center}
\begin{tabular}{|c|c|p{0.3\linewidth}|p{0.3\linewidth}|}
\hline 
\textbf{Metric} & \textbf{Type} & \textbf{Pros} & \textbf{Cons} \\ 
\hline 
Cores & Scale & Simple measure of number of cores deployed & Assumes equivalence between CPUs and GPUs; Reflects poorly on GPU codes; HPC facilities have far fewer GPUs than CPUs \\ 
\hline 
Threads & Scale & Gives strong indication of total number of parallel processes utilised by a job & Assumes a GPU thread and a CPU core are equivalent; Reflects poorly on CPU codes \\ 
\hline 
Nodes & Scale & Simple measure of total nodes used by a job & Resources available on a node varies widely between HPC facilities; GPUs may not be available on all nodes; Job may not demand all resources of a node \\ 
\hline 
Wall Time & Performance & Easiest measure to record & 	Geometry dependent measure \\ 
\hline 
Speed Up & Performance & Straightforward measure to interpret performance  & Derived unit from wall time or MLUPS \\ 
\hline 
MLUPS & Performance & More independent of geometry & Measure that is most relevant to LBM codes \\ 
\hline 
\end{tabular} 
\end{center}
\label{tab:PerformanceMetrics}
\end{table}

A particular challenge in comparing the performance GPU and CPU codes is the determination of an equivalence between the two architectures. By way of precedent, the Top500 \cite{Top500} list of supercomputers gives a measure of total `cores' in a supercomputer. For machines accelerated by GPUs, it appears that the GPU subunit of a `streaming multiprocessor' (in NVIDIA nomenclature) is deemed to be equivalent to a CPU core. By this measure, a NVIDIA V100 GPU corresponds to 80 CPU cores, whilst an A100 GPU is equivalent to 108. \\

Through association with a number of research projects, HemeLB-CPU has been able to be executed on a range of HPC machines and architectures including:
\begin{itemize}
\item Blue Waters (NCSA - 22,636 nodes, 16 CPU cores on each node)
\item ARCHER (EPCC - 4920 nodes, 24 CPU cores on each node)
\item ARCHER2 (EPCC - 5848 nodes (Phase 1 = 1024 nodes), 128 CPU cores on each node)
\item SuperMUC-NG (LRZ - 6480 nodes, 48 CPU cores on each node)
\end{itemize}
while HemeLB-GPU has been able to be executed on the following: 
\begin{itemize}
\item Piz-Daint (CSCS - 5704 nodes, 12 CPU cores and  1  P100  GPU  on  each node),
\item JUWELS-Cluster  (JSC - 56  nodes,  40  CPU cores  and 4 V100 GPUs on each node),
\item JUWELS-Booster (JSC - 936 nodes, 48 CPU cores and 4 A100 GPUs on each node) and
\item SUMMIT (ORNL – 4608 nodes, 42 CPU cores and 6 V100 GPUs on each node).
\end{itemize}

This breadth of machines has enabled us to develop a broad understanding of the HemeLB code and how it performs on a range of architectures. In our following analysis we will restrict reporting to cases where common test cases have been conducted. \\

\subsection{HemeLB-CPU - Strong scaling performance} 
\begin{figure}[!ht]
   \centering
	\includegraphics[width=0.98\textwidth]{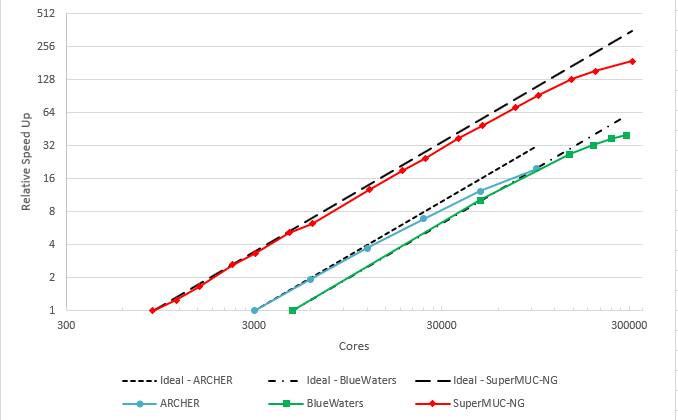}
   \caption{Strong scaling performance of the CPU code on various generations of CPU based machines}   \label{ScalingCPU}
\end{figure}

As the capabilities of both HemeLB and supercomputers have increased, the strong scaling performance of HemeLB has been repeatedly demonstrated, from tens to hundreds of thousands of compute cores. Here we highlight this improvement on HPC facilities of varying architecture and scale. On the SuperMUC-NG machine in particular, we had the opportunity to test the performance of HemeLB on a new supercomputer that was then within the top 10 of the Top500 list and represented one of the closest estimates of performance on an exascale machine available. Independently and in collaboration with the POP Centre of Excellence \cite{POPCoE}, we were able to conduct a similar test regime on Blue Waters and SuperMUC-NG using the circle of Willis geometry - an arterial structure typical of production jobs. The details of these studies are more fully reported in \cite{McCullough2021,Wylie2018,Wylie2020}. In these studies, we were able to run simulations at up to near-full or full-machine scale on both machines (Blue Waters 288,000 cores, 80\% of available cores; SuperMUC-NG 309,696 cores, 99.6\% of available cores). We report the performance of these in Figure \ref{ScalingCPU}. In Figure \ref{ScalingCPU}, we also provide comparison to a similar test conducted on ARCHER with a smaller circle of Willis test domain of 777 million lattice sites. The improvement in performance through the use of a geometry that can better occupy the compute capacity of a machine is clear. We observed 75\% strong scaling efficiency at 48,000 cores on ARCHER, 81\% efficiency at 192,000 cores on Blue Waters and 75\% efficiency on SuperMUC-NG at 147,456 cores. On all machines, HemeLB continues to scale strongly at all higher core counts tested. We believe that the tapering of performance at higher core counts can be attributed to two key reasons: 1) the test geometry not being large enough to fully occupy the compute cores at full machine scale; and 2) the impact of under-performing machine hardware or software components. These can impact the ability of machines to run large models effectively and can cause them to be unable to handle and display data generated from such simulations. Item 1) draws attention to an important issue which is only encountered by applications that seek to run models which require the full production partition in order to run: very often, the machines themselves may have not been optimised to routinely handle jobs that operate at this scale. This may mean that internode communication may take longer periods when large numbers of cores are being deployed. This makes it more difficult for computation to mask communication and good strong scaling performance to be achieved. Regarding item 2), on SuperMUC-NG, working with collaborators from POP CoE we identified several benchmark tests that exhibited significantly reduced performance due to a single faulty compute node. Even with such nodes excluded, the use of these extremely large core counts exposes benchmark tests to vagaries in performance that are frequently difficult and/or expensive to quantify. \\

\subsection{HemeLB-GPU - Strong Scaling performance}
\begin{figure}[!ht]
\includegraphics[width=0.98\textwidth]{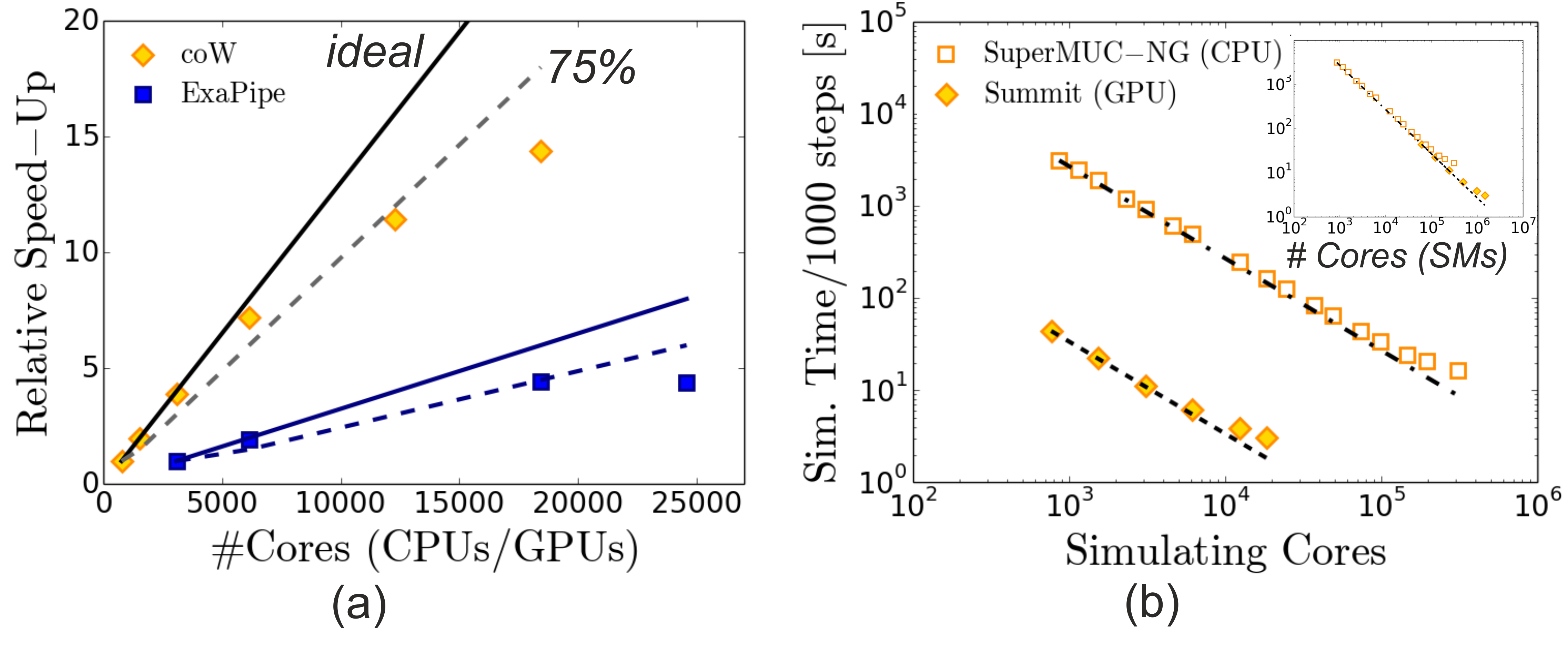}
   \caption{Strong scaling performance of HemeLB: (a) Relative speed-up of HemeLB-GPU on Summit. (b) Comparison of strong scaling performance of HemeLB-CPU and HemeLB-GPU codes using a common geometry (circle of Willis). The normalised walltime recorded by each architecture is plotted against the raw number of resources (CPU/GPU cores) requested. Inset: Normalised walltime as a function of CPU equivalent hardware units (CPU cores and SMs). }
   \label{fig:Scaling_Compare_GPU_CPU}
\end{figure}

We have been able to test the performance of the GPU code on Summit – the second fastest machine on the current Top500 list {Top500} whose performance is accelerated by 27,648 NVIDIA V100 GPUs. 
Using the circle of Willis geometry (Fig.~\ref{fig:TestDomains}(b)) and comparing to a base measurement on 768 GPU cores, Fig.~\ref{fig:Scaling_Compare_GPU_CPU}(a) demonstrates 90\% perfect scaling performance on 6,144 GPUs and continued strong scaling to 18,432 GPUs, which is approximately 2/3 of Summit’s capacity. The strong scaling efficiency drops to 72\% at 12,288 GPUs and 60\% at 18,432 GPUs. 
In an extra test on Summit, we examined the large-scale performance by increasing the simulation domain and examining the scaling characteristics of a cylinder constructed of $\sim 37.5\times 10^9$  sites, labeled as ExaPipe geometry. As shown in Figure \ref{fig:Scaling_Compare_GPU_CPU}(a), this improves the strong scaling performance and leads to 74\% strong scaling efficiency at 18,432  GPUs, due to the increased computation to communication ratio. 
This  makes  evident  the improvement in performance through the use of a geometry that can better occupy the compute capacity of the machine. \\

To enable comparison of the HemeLB-CPU and HemeLB-GPU codes we utilised the same circle of Willis geometry as on Blue Waters and SuperMUC-NG. 
Figure ~\ref{fig:Scaling_Compare_GPU_CPU}(b) provides a side-by-side comparison of the SuperMUC-NG and Summit results using an equivalence metric of 1 CPU core to 1 GPU. Here we can observe that the acceleration of the GPU code for an equivalent number of `cores' leads to a speed-up factor of approximately $\times 85$. For both codes, the tapering of results at the highest core counts is reflective of the limitation of the size of the geometry at these scales. \\ 

Due to the fundamental nature of GPU operation, a measure of cores is not necessarily the most relevant measure of parallelism in such applications. In the inset of Figure \ref{fig:Scaling_Compare_GPU_CPU}(b) we compare the performance of the CPU and GPU codes using the equivalence suggested by the Top500 list, where NVIDIA V100 GPUs are equivalent to 80 CPU cores. Probably this arises from the fact that each Tesla V100 features 80 streaming multiprocessors (SMs).
Here, it can be seen how the GPU code is able to continue the scalability characteristics to a higher level of parallelism than is achieved with the CPU codes. The total level of parallelism exhibited by the GPU code can be estimated by taking into account the fact that the NVIDIA V100 GPUs available on Summit each possess 5120 FP64 CUDA cores that execute tasks in parallel. As such, the performance we report on Summit represents 90\% efficient scaling on over 31 million parallel tasks and continued strong scaling behaviour up to over 94 million processes. This helps to support our argument that the inherent parallelism of HemeLB is well positioned to be deployed on an exascale machine. \\

\begin{figure}[!ht]
\includegraphics[width=0.98\textwidth]{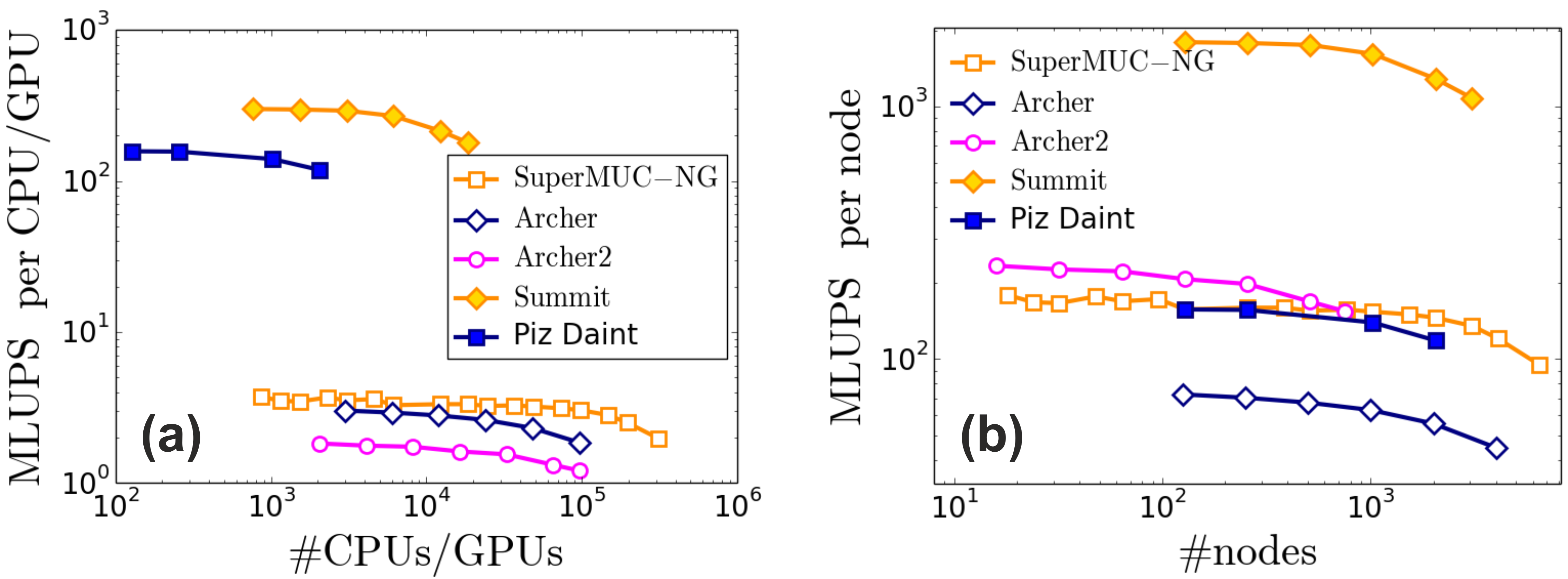}
   \caption{Performance of HemeLB reported in millions of lattice sites per second (MLUPS) on (a) a per core basis, MLUPSpc, and (b) per node basis MLUPSpn. Open symbols correspond to results from CPU-only HPC machines (HemeLB-CPU), while the ones with filled symbols from HPC machines with NVIDIA GPUs (HemeLB-GPU). 
 Simulations were	 conducted using a range of domains: small circle of Willis - ARCHER; large circle of Willis - Blue Waters, SuperMUC-NG, Summit; Full human veins - ARCHER, ARCHER2, Piz Daint. }
   \label{fig:MLUPS_per_cores_nodes}
\end{figure}

In conjunction with the scaling tests discussed above, we have also generated some more normalised performance data for the CPU and GPU versions of HemeLB on a wider variety of machines. 
Figure \ref{fig:MLUPS_per_cores_nodes} reports the performance of simulations in Millions of Lattice site Updates Per Second (MLUPS) on a per core and node basis. 
MLUPS is defined as 
\begin{equation}
{\rm MLUPS}=\frac{nFluidSites\times nTimeSteps }{SimTime\times10^6}\,
\end{equation}
where $nFluidSites$, $nTimeSteps$ and $SimTime$ are the total number of fluid sites in the simulation domain, number of time-steps and total simulation time in seconds respectively. The evaluation of the performance on a per core (MLUPSpc) and per node (MLUPSpn) basis is based on the following definitions
\begin{equation}
{\rm MLUPSpc}=\frac{\rm MLUPS}{nCores} 
\quad\mathrm{and}\quad 
{\rm MLUPSpn}=\frac{\rm MLUPS}{nNodes}
\end{equation}
where $nCores$, $nNodes$ are the total number of CPU cores (MPI ranks) and  total compute nodes respectively. Here, we must note that we used one MPI rank per GPU, although  HemeLB-GPU can also run using other configurations as well by having multiple MPI ranks accessing the same GPU.

The above is informative as we have access to a broad range of high-end machines with different configurations of CPUs, GPUs, memory and bandwidth. For all machines, the drop in performance at higher node counts is due to the test domain not being sufficiently large to occupy the resources at that scale, the point at which this occurs varrying between machines due to the different memory availability on CPU cores and GPU cards. Here we can see how the construction of a node can impact performance. For example, Piz Daint, a machine with only a single NVIDIA P100 GPU per node, delivers comparable performance to a single, CPU only, node on SuperMUC-NG. By comparison, the nodes on Summit, containing 6 NVIDIA V100s, deliver a performance improvement of a factor of at least 10 on other reported machines – i.e. 1 Summit node delivers the same performance as 10 SuperMUC-NG nodes.  This indicates that the largest data for Summit reported in Figure \ref{fig:Scaling_Compare_GPU_CPU} corresponds to over 30,000 SuperMUC-NG nodes – 5 times larger than the actual SuperMUC-NG. \\ 

The comparison between the 2013-era ARCHER machine and the current generation ARCHER2 provides another notable comparison. Here, with more than 5 times as many cores per node, the new machine can only generate a node-based performance improvement of a factor of 2.9. This is due to hardware decisions that were made during procurement that led to a net reduction in memory bandwidth per core. Such decisions can have a negative impact on the relative performance of monolithic codes, especially those that are memory bound such as the lattice Boltzmann method. However, the same decisions can also be advantageous for other compute patterns that are less dependent on such hardware choices. The original ARCHER machine possessed 2$\times$12-core Intel Ivy Bridge E5-2697 v2 CPUs on each node whilst the newer ARCHER2 has 2$\times$64-core AMD Zen2 7742 CPUs per node. Whilst the AMD CPUs do run at a lower frequency than the Intel cores, this is not sufficient to fully explain the per-core difference in performance between the two machines. On earlier generations of Intel and AMD hardware, a similar difference in performance for a lattice Boltzmann code was reported in \cite{Schoenherr2011}.\\

The variation in performance characteristics between the core and node based performance in Figure \ref{fig:MLUPS_per_cores_nodes} highlights the challenges posed when designing HPC architectures. Choices such as the number of GPUs and CPUs deployed on a node, memory availability and network will all have an impact on code performance and will vary depending different codes' operational requirements. Equally, different compute patterns will also pose competing demands on HPC performance. \\

\section{Challenges at the emerging exascale}
\begin{figure}[!ht]
 \begin{subfigure}{0.7\textwidth}
   \centering
   \includegraphics[width=0.98\textwidth]{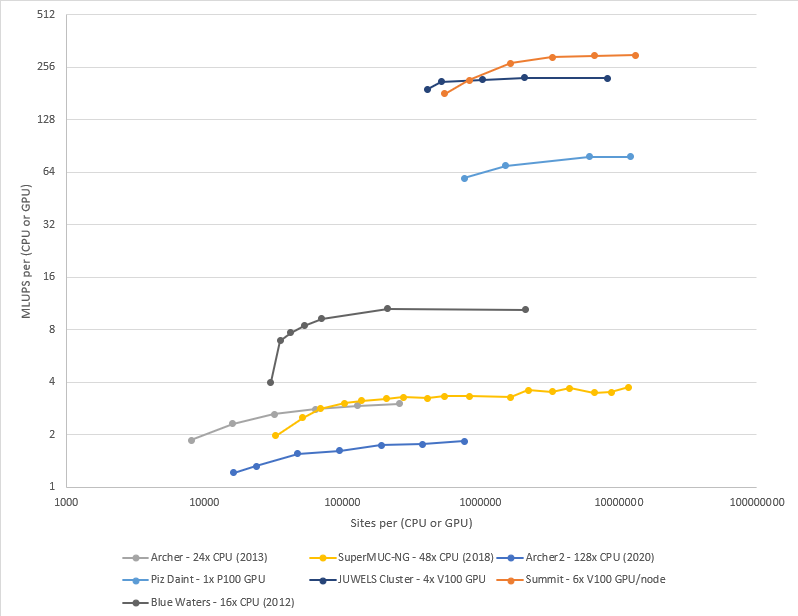}
   \caption{MLUPS vs sites per hardware unit}
   \label{MLUPScores}
 \end{subfigure}
 \begin{subfigure}{0.7\textwidth}
   \centering
   \includegraphics[width=0.98\textwidth]{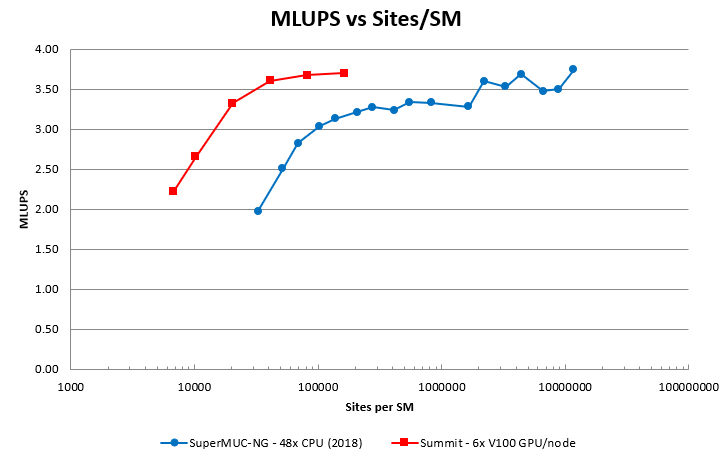}
   \caption{MLUPS vs sites per CPU core equivalent on SuperMUC-NG and Summit}
   \label{MLUPSsites}
 \end{subfigure}
  \caption{One challenge of achieving effective exascale simulation will be to generate geometries that are large enough to effectively occupy the hardware. Here we illustrate the drop in compute performance that is observed at various values of sites per core on both CPU and GPU architectures. The point at which this occurs varies between machines.}
  \label{fig:CorePerformance}
\end{figure}

Based on our current investigations, the compute and thus scaling performance of HemeLB is dependent on the number of lattice sites per core. This is proportional to the computational time per LB iteration; in the case that this is sufficient to hide the MPI communication, then the strong scaling behavior can be maintained. 
This is demonstrated in Figure \ref{fig:CorePerformance}, where we plot the simulation speed against the number of sites per core for the CPU code. A significant performance drop is seen below $\sim10^5$ sites/core for HemeLB-CPU and $\sim10^6$ sites/core for HemeLB-GPU, demonstrating this to be a limit at which computation is outweighed by communication between nodes. 

We believe that a full human vascular model may require at least $50\times10^9$ lattice sites to resolve the model at high resolution. Such a geometry, being five times larger than that used in the POP evaluation, will enable many of the performance concerns seen at scale on SuperMUC-NG to be addressed. The challenges associated with creating, storing and simulating a geometry of this scale are highly non-trivial. Regardless of the processing method chosen, storage requirements can be on the order of tens of terabytes for interim files and processing time may exceed 50 hours, both of which are factors that may exceed allocation fair-use limitations imposed by HPC facilities. For future exascale machines with some 10 million cores, the potential requirement of 10$^{12}$ lattice sites to achieve good scaling will further challenge the mesh generation of HemeLB (and indeed any code) and the machine’s storage infrastructure. The HemeLB developers remain in collaboration with operators of SuperMUC-NG to develop solutions for addressing these concerns.

\begin{table}
\caption{Minimum memory demands for large-scale geometry generation in HemeLB on a CPU only machine, all figures in TB.}
\begin{center}
\scalebox{0.9}{
\begin{tabular}{|c|p{3cm}|c|c|c|}
\hline 
\textbf{Cores} & \textbf{Classification} & \textbf{10$^5$ sites/core} & \textbf{5x10$^5$ sites/core} & \textbf{10$^6$ sites/core} \\ 
\hline 
10$^5$ & Current HPC - medium scale & 12.72 & 63.6 & 127.2 \\ 
\hline 
5x10$^5$ & Current HPC - largest scale & 63.6 & 318 & 636 \\ 
\hline 
10$^6$ & Next-Generation HPC & 127.2 & 636 & 1272 \\ 
\hline 
10$^7$ & Exascale HPC & 1272 & 6360 & 12720 \\ 
\hline 
\end{tabular} 
}
\end{center}
\end{table}

\begin{figure}[!ht]
\includegraphics[width=0.98\textwidth]{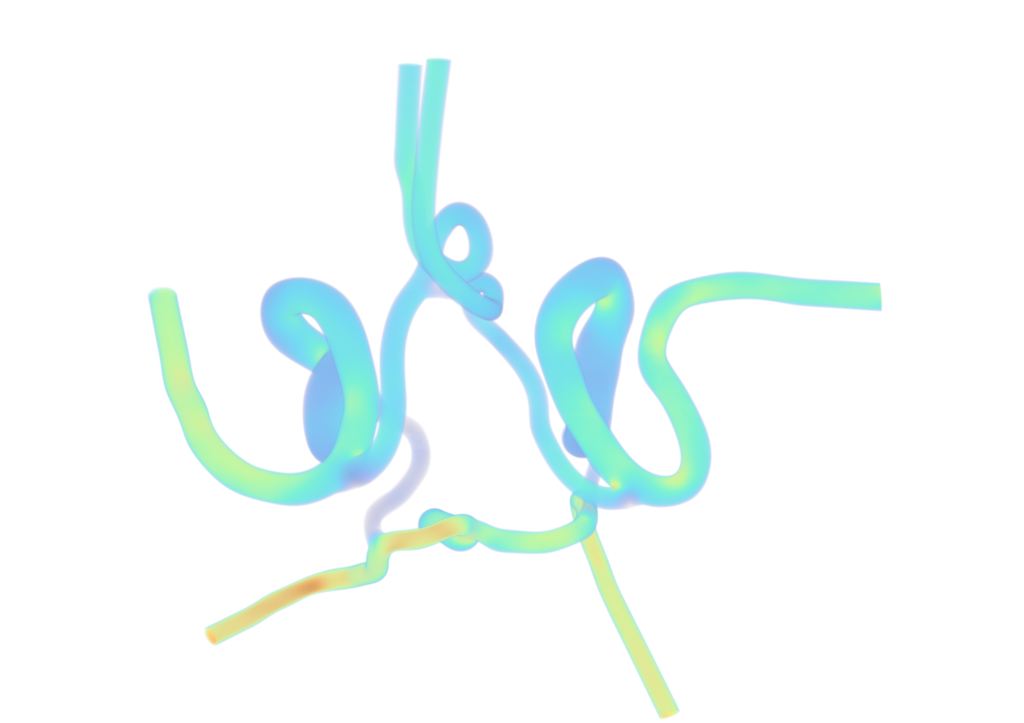}
   \caption{Example visualisation of a velocity field within the circle of Willis domain using the workflow developed in a collaboration with LRZ and Intel. This demonstrates how high performance computers can be deployed to efficiently render images of the very large data sets created by HemeLB with human-scale data. As the data structures output by the CPU and GPU versions of the code is the same this tool can be used with both.
}
   \label{fig:LRZvizExample}
\end{figure}

Whilst the figures presented here are particular to our deployments of HemeLB, they remain illustrative of the challenges of migrating monolithic codes to exascale HPC facilities. In addition to the storage challenges related to generating geometries of sufficient scale, the operational demands of loading such a domain for simulation and writing and storing output will put further demands on the resources of an exascale machine. The time required for pre- and post-processing operations generally combine to frame the challenge of the overall efficiency of operating at scale. This is particularly related to how quickly a useful result can be obtained from new initial data - in the case of HemeLB this may be an stl file of a patient's vascular structure. The scaling data presented here focuses on the simulation phase of the code. Demands of data writing and post-processing in particular may be harder to predict in general as the output requirements can be specific to a particular problem. The tools used for pre- and post-processing may not have received the same degree of performance optimization as the simulation kernels and highlight the need to consider the whole simulation workflow when evaluating the performance of a code. The use of workflow management tools such as RADICAL CyberTools \cite{RadicalCT} can help to manage multiple processes concurrently to help reduce resource wastage and increase efficiency; this is particularly true for ensemble type simulation studies. For the post-processing of HemeLB data an ongoing collaboration with the LRZ Centre for Virtual Reality and Visualisation and Intel has developed a workflow based on Intel OSPRay Studio to enable rapid visualisation of the very large and complex datasets that we can routinely generate \cite{HemeLBVIZ_ISC21}. Running this on HPC systems as part of a computational workflow enables an immersive rendering of vascular data to be obtained. Future work in this direction envisions the development of a 3D virtual reality experience of simulation data and the capacity for computational steering of a running simulation. As the output data structures generated by the CPU and GPU codes are the same, this tool is compatible with both versions of the code. An example of the results that can be achieved with this workflow is demonstrated for the circle of Willis case study in Figure \ref{fig:LRZvizExample}. \\

The ability for a given application to achieve optimum performance on a particular machine will require an ongoing co-design effort between both code developers, HPC operators and hardware experts. As the specific choice of hardware and its deployment can have significant impacts on the performance of a code, it is incumbent upon operators to ensure that the choices that they make during procurement bring the greatest benefit to the widest cross-section of their users of all application types. There is an equal onus on code developers to be looking forward to optimisations and performance gains that can be obtained on a given piece of hardware such as through compiler options as new options become available, or through an update of the code itself. These demands must also be balanced against the need to generate scientific knowledge from a code, where the ability to scale efficiently to large resource counts may not be the most decisive criteria.  \\

\section{Blood flow simulations with the HemeLB-GPU code}
\begin{figure}[!ht]
\includegraphics[width=0.98\textwidth]{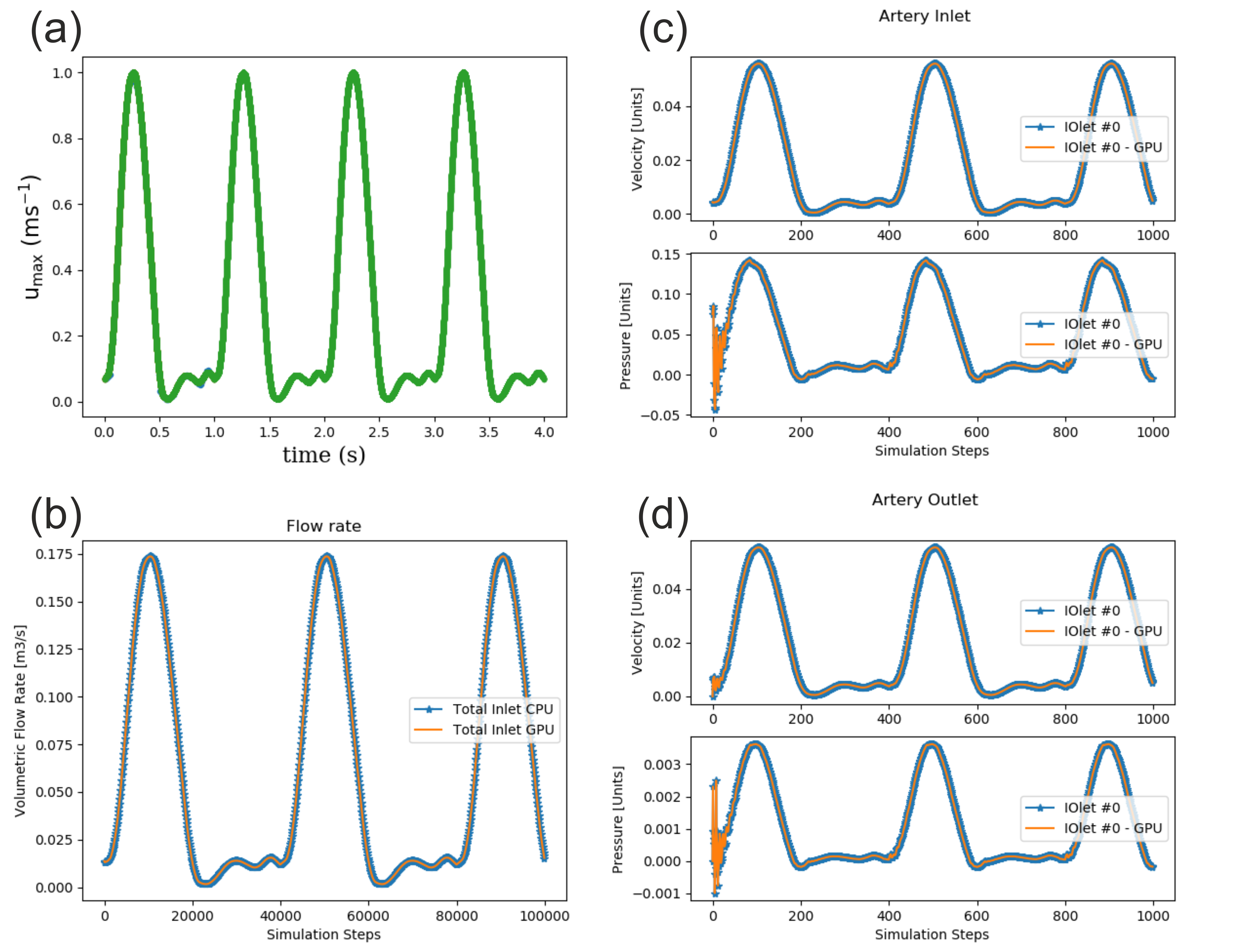}
   \caption{Comparison of HemeLB-CPU and HemeLB-GPU codes. Simulation results using a pipe domain with velocity BCs imposed at the inlet and fixed constant pressure at the outlet.
   (a) The imposed maximum velocity profile imposed at the inlet corresponding to a heart beat profile of 60 beats per minute. 
   (b) Comparison of the volumetric flow rates.
   (c) Comparison of maximum velocity and pressure at the inlet.
   (d) Comparison of maximum velocity and pressure at the outlet.  
}
   \label{fig:compare_CPU_GPU_pipe_VelPres}
\end{figure}

Here we provide a simple example to demonstrate HemeLB-GPU code's capacity of running blood flow simulations. We present results from a simulation using a simple pipe domain. In this simulation we apply the indicated velocity profile to the inlet, see Fig.~\ref{fig:compare_CPU_GPU_pipe_VelPres}(a), with a beat profile of 60 beats per minute. The profile shown indicates the maximum velocity within the inlet; this is scaled by the weighting of the specific inlet profile, i.e. each inlet point uses an assigned weight based on its position within the cross-sectional plane. A fixed constant pressure is set at the outlet boundary. We compare the simulation results produced by HemeLB-CPU and HemeLB-GPU; Figs.~\ref{fig:compare_CPU_GPU_pipe_VelPres}(b)-(d) demonstrate that we get identical results for the 2 versions of the code.

\section{Future Work}\label{sec:Future_plans}
We outlined here a first implementation of a GPU accelerated version of HemeLB using CUDA C++. 
Results demonstrate that the code continues to exhibit excellent strong scaling performance up to thousands of GPUs and a decent speed up compared to HemeLB-CPU. This is an excellent outcome; however, code development and performance optimisation is an ongoing process, considering that single phase LB codes have demonstrated an even further gain in performance in terms of MLUPS per core \cite{hemelb_2018}. 
Future work could focus on various issues, including the following:
\begin{enumerate}
\item Optimise the GPU kernels' performance. Tools and metrics provided by NVIDIA Nsight Compute and Nsight Systems can help to this end, for example the register usage and the GPU occupancy. The available registers per streaming multiprocessor (SM) imposes a restriction on the number of active warps on the SM and consequently may impact the GPU occupancy, leading to performance degradation. Currently the maximum registers usage of the GPU kernels is around 75, which can be reduced significantly. 
\item Improve boundary conditions at iolets. Currently HemeLB-GPU supports driving the fluid flow by using either pressure or velocity boundary conditions, which can vary as a function of time. More physiologically correct boundary conditions should be applied though, when running patient- specific flow simulations, especially for outlet BCs to account for downstream resistance \cite{zhang2014numerical, maurits2007influence, feiger2020accelerating}. However, these resistance models require feedback from experiments and iterative tuning at each outlet, so that eventually the simulated and \textit{in vivo} volumetric flow rates can match. 
\item Convert from the Push to the Pull-scheme \cite{wellein2006single,tran2017performance}. This refers to modifying the sequence of the streaming and collision steps of the LB algorithm. The Push-scheme, currently implemented, refers to the situation when the collision precedes the streaming step, while the reversed situation is known as the Pull-scheme. 
The fundamental difference of the 2 schemes lies with the ordering of uncoalesced memory accesses during reading from or writing to the GPU global memory.
It was shown that the Pull scheme performs better than the Push-scheme, due to the cost of uncoalesced memory accesses during reading from the GPU device memory (Pull-scheme) being lower than during writing (Push-scheme) \cite{wellein2006single,tran2017performance}.
\item Implement elastic walls to better represent realistic vascular geometries, following our recent implementation in HemeLB-CPU, which most importantly demonstrated that this is feasible without a loss of computational performance \cite{McCullough2022_EW}. 
\end{enumerate}

As exascale HPC becomes available, these improvements, in conjunction with the capability described in this paper, will continue to ensure that HemeLB is able to take full advantage of these machines for deployment in human-scale blood flow simulations. In an effort to reach this goal for use in the field of personalised medicine, we have developed a self-coupled version of HemeLB that allows simultaneous study of arterial and venous domains \cite{McCullough2021} and provided an illustration of how it can be used in problems of clinical interest \cite{McCullough2022_AVF}. The enhanced performance capabilities of the GPU code will enable these CPU implementations to be conducted more quickly whilst retaining HemeLB's known strong scaling characteristics. \\

\section{Conclusions}\label{sec:Conclusions}
In this paper we have outlined a version of HemeLB capable of execution on NVIDIA GPUs that delivers excellent strong scaling performance to very large numbers of GPUs. This has been built off a CPU-only version of HemeLB that has demonstrated similar strong scaling performance to the full production partition of the German supercomputer SuperMUC-NG.  We have examined the new code's performance on a number of difference HPC platforms that represent a broad spectrum of hardware manufacturers and deployment frameworks. We illustrate that the performance of a code can be highly dependent on these factors with newer configurations not necessarily yielding better performance. \\

As the arrival of exascale machines moves closer, we will continue to develop all versions of HemeLB to aim for best performance on the widest range of machines. For example, the CPU code will look to incorporate the latest optimizations for code compilation and performance. Similarly we will continue to optimize the GPU port and aim to convert from our current CUDA-based implementation to one that allows deployment on both NVIDIA and AMD GPUs. This second objective will allow us to run HemeLB on a wider range of GPU accelerated HPC. As new features and libraries become available, we anticipate that further optimizations can be made in the GPU code to further reduce communication time at scale and further improve the scaling characteristics of our tool. \\ 

The results that we have presented in this paper highlight the role that all stakeholders have in determining the specifications of future HPC facilities. This can be regarded as a co-design exercise that will be beneficial to both code developers and HPC operators alike. Achieving a balance between the hardware demands of different compute patterns will be critical to generating the best outcomes for a wide range of scientific applications. Making future users of a new HPC facility aware of the hardware design as early as possible will let code developers adapt to any changes that may be necessary. 

\section*{Funding}
We acknowledge funding support from European Commission CompBioMed Centre of Excellence (Grant No. 675451 and 823712). Support from the UK Engineering and Physical Sciences Research Council under the project `UK Consortium on Mesoscale Engineering Sciences (UKCOMES)' (Grant No. EP/R029598/1) is gratefully acknowledged. We acknowledge funding support from MRC for a Medical Bioinformatics grant (MR/L016311/1), and special funding from the UCL Provost. \\

The authors gratefully acknowledge the Gauss Centre for Supercomputing e.V. (\url{www.gauss-centre.eu}) for funding this project by providing computing time on the GCS Supercomputer SuperMUC-NG at Leibniz Supercomputing Centre (\url{www.lrz.de}). This work used the ARCHER UK National Supercomputing Service (http://www.archer.ac.uk) as well as the ARCHER2 UK National Supercomputing Service (http://www.archer2.ac.uk). We acknowledge PRACE for awarding us access to JUWELS at GCS@FZJ, Germany and Piz Daint at CSCS, Switzerland. This research used resources of the Oak Ridge Leadership Computing Facility at the Oak Ridge National Laboratory, which is supported by the Office of Science of the U.S. Department of Energy under Contract No. DE-AC05-00OR22725.



\bibliographystyle{unsrtnat_JMmod}
\bibliography{2021_HemeLBGPUPerformancePaper}

\end{document}